\documentclass[fleqn,usenatbib]{mnras}

\usepackage{newtxtext,newtxmath}
\usepackage{multirow}

\usepackage[T1]{fontenc}
\usepackage{ae,aecompl}

\usepackage{graphicx}	
\usepackage{amsmath}	

\newcommand{\tabhead}[1]{\textbf{#1}}

\newcommand{\comment}[1]{}

\usepackage[flushleft]{threeparttable}

\newcommand{\targ}{XY\,Ari}
\newcommand{\kms}{ km~s$^{-1}$}

\title[Stellar masses in XY Ari]{Dynamical mass of the white dwarf in XY Ari: a test for intermediate polar X-ray spectral models }

\author[A. \'Alvarez-Hern\'andez et al.]
       {A. \'Alvarez-Hern\'andez$^{1,2}$\thanks{E-mail: ayozeav@iac.es}, 
        M. A. P. Torres$^{1,2}$, P. Rodr\'\i guez-Gil$^{1,2}$, T. Shahbaz$^{1,2}$, J. S\'anchez-Sierras$^{1,2}$,  \newauthor J. A. Acosta-Pulido$^{1,2}$, P. G. Jonker$^{3,4}$, K. D. Gazeas$^{5}$, P. Hakala$^{6}$, J. M. Corral-Santana$^{7}$\\
\\
	$^1$Instituto de Astrof\'\i sica de Canarias, E-38205 La Laguna, Tenerife, Spain\\
        $^2$Departamento de Astrof\'\i sica, Universidad de La Laguna, E-38206 La Laguna, Tenerife, Spain\\    
        $^{3}$SRON, Netherlands Institute for Space Research, Niels Bohrweg 4, 2333~CA, Leiden, The Netherlands\\
        $^{4}$Department of Astrophysics/ IMAPP, Radboud University, Heyendaalseweg 135,6525 AJ, Nijmegen, The Netherlands\\
        $^5$Section of Astrophysics, Astronomy and Mechanics, Department of Physics, National and Kapodistrian University of Athens, GR-15784 Zografos, Athens, Greece\\
        $^{6}$Finnish Centre for Astronomy with ESO (FINCA), Quantum, Vesilinnantie 5, FI-20014 University of Turku, Finland\\
        $^{7}$European Southern Observatory, Alonso de C\'ordova 3107, Vitacura, Casilla 19001, Santiago de Chile, Chile\\}

\date{Accepted XXX. Received YYY; in original form ZZZ}

\pubyear{2023}

\begin{document}
\label{firstpage}
\pagerange{\pageref{firstpage}--\pageref{lastpage}}
\maketitle

\begin{abstract}
We present a dynamical study of the eclipsing intermediate polar \targ\ based on time-resolved near-infrared spectroscopy obtained with the EMIR spectrograph on the 10.4-m Gran Telescopio Canarias. Using main sequence template spectra taken with the same instrument setup as the target spectra, we measure a radial velocity amplitude of the late K-type donor star $K_2=256 \pm 2$~\kms. We also obtain the rotational broadening of its photospheric lines $v_\mathrm{rot} \sin  i = 141 \pm 3$~\kms . From these and the eclipse geometry, we derive a donor-to-white-dwarf mass ratio $q = M_2/M_1 = 0.62 \pm 0.02$, an orbital inclination $i = 80.8^{\circ} \pm 0.5^{\circ}$ and dynamical masses $M_{1} = 1.21 \pm 0.04 \, \mathrm{M}_{\odot}$ and $M_2 = 0.75 \pm 0.04 \, \mathrm{M}_{\odot}$ ($1 \sigma$). This result places the white dwarf in XY Ari as one of the three most massive known in a cataclysmic variable. Comparison with white dwarf mass estimates from X-ray spectral studies could indicate the necessity of an improvement of the X-ray models and/or analysis techniques, as a number of X-ray white dwarf masses are in disagreement with the dynamical mass value. 
\end{abstract}




\section{Introduction} \label{intro}

Cataclysmic variables (CVs) are binary systems typically consisting of a non-degenerate donor star that fills its Roche lobe and transfers mass to a white dwarf (WD) primary star \citep[see \citealt{warner-libro} for a review]{kraft-64}. In the subclass of CVs known as intermediate polars (IPs), the WD has a relatively intense magnetic field ($10^4-10^6$~G), which is not strong enough to disrupt the accretion disc entirely (as in the case of the polar CVs). Instead, the disc is truncated at some radius from the WD and the accretion flow is conducted from there to the WD magnetic poles following the field lines.

\targ\ is an IP first discovered by the \textit{Einstein} satellite as an X-ray source (1H 0253+193) coincident with the centre of the molecular cloud Lynds 1457 \citep{halpern-87}. For this reason, it was initially suggested as a T Tauri star embedded in the cloud. Observations performed with the \textit{Ginga} satellite discovered coherent pulsations with a period of $\simeq 206$~s \citep{takano-89,koyama-91}. Based on this, \cite{patterson-90} rejected the possibility of 1H 0253+193 being an X-ray binary and proposed either an isolated WD or neutron star accreting gas from the cloud or a CV. Later, \cite{kamata-91} discovered periodic dips in the X-ray light curve with a period of 21829~s~$\simeq6.06$~h, which were interpreted as eclipses of the X-ray source produced by the transit of a donor star. They concluded that 1H 0253+193 is most likely a CV located behind Lynds 1457. \cite{zuckerman-92} identified the near-infrared (NIR) counterpart at $J \simeq 16.0$, $H \simeq 14.2$ and $K \simeq 13.3$. Similar magnitudes were reported in the Two Micron All Sky Survey (2MASS; \citealt{2mass}). All these values correspond to the quiescent state of the system, since \targ\ is known to undergo dwarf nova outbursts \citep{hellier-97-outburst}. Due to the high extinction produced by the molecular cloud, \targ\ is only fully detected in the redder bandpasses of the Sloan Digital Sky Survey (SDSS) and Panoramic Survey Telescope and Rapid Response System (Pan-STARRS) surveys at $i = 22 - 21.5$, $z = 20.0 - 19.4$ \citep{sdss,ps1}.

\begin{table*}
\caption[]{Log of the \targ\ GTC/EMIR spectroscopy. Orbital phases are calculated using the ephemeris derived in Section~\ref{sec-k2}. The on-source time was 480~s for each ABBA cycle. The seeing values correspond to the FWHM  of the spatial profile at spectral positions near $\lambda = 2.2~\mu\mathrm{m}$. Magnitudes for XY~Ari were measured in the acquisition images by performing differential photometry relative to the nearby star 2MASS 02561213+1925362.}
\label{tab:observaciones}
\centering
\begin{tabular}{l c c c c c}
\hline\noalign{\smallskip}
Date  & ABBA & Orbital phase & Airmass & Seeing & $K_\mathrm{s}$\\
\textbf{}  & cycles & coverage & \textbf{} & (arcsec) & (mag) \\
\hline\noalign{\smallskip}
2018 Dec 27 & 4 & 0.63--0.71 & 1.18--1.26 & 0.7--0.9 & 13.5--14.0\\ 
2020 Jan 05 & 4 & 0.15--0.23 & 1.01--1.02 & 0.9--1.0 & 13.1--13.2\\ 
2020 Sep 28 & 4 & 0.68--0.75 & 1.02--1.05 & 0.9--1.1 & 13.5\\
2020 Sep 30 & 8 & 0.92--0.99, 0.12--0.20 & 1.04--1.22 & 0.7--0.8 & 13.2--13.9\\ 
2020 Oct 01 & 8 & 0.79--0.86, 0.98--0.06 & 1.01--1.14 & 0.6--0.7 & 13.4--13.9\\ 
2020 Nov 23 & 8 & 0.25--0.32, 0.49--0.56 & 1.09--1.50 & 1.0--0.8 & 13.2--13.5\\ 
2022 Feb 11 & 4 & 0.50--0.57 & 1.48--1.67 & 1.0--1.3 & $^*$\\ 
\hline\noalign{\smallskip}
\multicolumn{4}{c}{\footnotesize{$^*$ Acquisition in the $J$-band. We measure $J=16.1-16.6$.}} \\
\end{tabular}
\end{table*}

Some of the orbital parameters of \targ\ have been robustly measured from X-ray and NIR data. \cite{allan-96} took $H$-band photometry covering more than three orbits of the system. Combining eclipse timings from their light curves and from \cite{kamata-91} and \cite{zuckerman-92}, they derived precise mid-eclipse ephemeris. They also modelled their $H$-band light curve, which is dominated by the ellipsoidal variation caused by the change of the projected surface of a non-irradiated Roche-lobe-filling donor star with orbital phase. From the modelling they set constraints of $0.43 < q < 0.71$ and $80^{\circ} < i < 87^{\circ}$ on the binary mass ratio and the orbital inclination, respectively. \cite{hellier-97} used X-ray data from the \textit{RXTE} satellite to study the accreting regions of the WD and found that accretion takes place on both magnetic poles, which can be seen in quiescence. From imposing that the lower accreting pole of the WD is seen through the inner hole in the accretion disc, he refined the previous ranges to $0.48 < q < 0.68$ and $80^{\circ} < i < 84^{\circ}$. Later, \cite{littlefair-2001} presented low-resolution $K$-band spectroscopy and applied the optimal subtraction technique \citep{marsh-94} to classify the donor star. Their results supported a K7 -- M0~V spectral type, most probable M0~$\pm$~0.5~V.

Dynamical masses of the stellar components in \targ\ have never been derived, since the radial velocity amplitude of the donor star has not been measured and the binary mass ratio is poorly constrained. Estimates of the WD mass from X-ray spectral modelling have been reported, but the results are highly inconsistent with each other, with masses in a wide range ($0.80 - 1.50 \, \mathrm{M}_{\odot}$ at $1 \sigma$). In this work, we present a complete dynamical study of \targ , obtaining for the first time robust values of the stellar masses. We compare our results with previous estimates from X-ray spectroscopy and provide valuable feedback to the X-ray spectral modelling. 

The article is structured as follows: in Section~\ref{sec-obs} we present our $K$-band spectroscopic observations and their reduction. From their analysis we obtain the radial velocity curve of the donor star (Section~\ref{sec-k2}), its spectral type (Section~\ref{sec-spectralclassification}) and rotational broadening of the absorption lines (Section~\ref{sec-rotationalbroadening}). We then derive the binary mass ratio, the orbital inclination and the dynamical masses (Section~\ref{sec-masses}). In Section~\ref{sec-discuss} we discuss our results and in Section~\ref{sec-conclusions} we present our conclusions.

\section{Observations and data reduction}\label{sec-obs}

We obtained NIR spectroscopy of XY Ari using the EMIR spectrograph \citep{garzon-16, garzon-22} attached to the 10.4-m Gran Telescopio Canarias at the Observatorio del Roque de los Muchachos on the island of La Palma, Spain. The target was observed in queue mode at different epochs between $2018 - 2022$, covering $\simeq 65$ per cent of its 6.06-h orbit. We used the $K$-grism and implemented with the Configurable Slit Unit a 0.6-arcsec wide long slit positioned one arcmin left from the centre of the field of view. This instrument setup covered the $2.08-2.43~\mu\mathrm{m}$ wavelength range with a dispersion of 1.71~\AA\,$\mathrm{pix}^{-1}$ and a resolution of $\simeq 4.8$~\AA\ full-width at half-maximum (FWHM), as measured from the atmospheric OH emission lines. This is equivalent to $\simeq 65$~\kms\ at $2.2~\mu\mathrm{m}$. Each observing visit (block) to the target consisted of four consecutive ABBA nodding cycles using a nod of 12~arcsec and 120-s individual exposures. In total, we obtained 40 ABBA cycles on seven different nights (see Table~\ref{tab:observaciones} for a log of the observations). Only the last night data were taken at an airmass larger than $1.5$.

We also took spectra of template stars with the same setup as for XY~Ari. We observed main sequence stars with effective temperatures in the range $3020-4500\,\mathrm{K}$, covering spectral types from K4 to M5, selected from the samples in \cite{rojas-2012} and \cite{yee-2017}. All these template stars have low rotational velocities. For telluric line removal, we took spectra of A0~V stars close in sky position and time to the science and template spectra. The image quality measured by fitting the spatial profiles of the XY Ari, template and telluric star spectra showed that all data were taken under slit-limited conditions (see Table~\ref{tab:observaciones}).

The data reduction was performed using version 0.17.0 of the {\sc pyemir} package \citep{cardiel-19}. After dark and flat-field correction, the wavelength calibration and rectification were performed over the 2D frames using the sky OH emission lines. The rms scatter of the fits was $0.5-0.8$~\AA\ (equivalent to $7-11$~\kms) for all data sets. The wavelength calibration was examined by checking the telluric absorption features in the XY Ari data relative to their average spectrum, finding minor shifts  with a standard deviation of $5$~\kms . The four spectra of each ABBA cycle were averaged, sky subtracted and dithering corrected to obtain one spectrum per nodding cycle. This spectrum was extracted with the \textit{apall} task in \textsc{iraf}\footnote{{\sc iraf} is distributed by the National Optical Astronomy Observatories.}.

Two different techniques for the removal of telluric absorption features were tested. The first technique derives the Earth's atmospheric transmission by modelling the intrinsic spectra of the A0~V stars that were obtained close to each XY~Ari observation. This was done with a modified version of the \textsc{xtellcor general} (\textsc{idl}) package \citep{vacca-2003}. The second technique computes the atmospheric transmission spectrum from fitting the  telluric absorptions in the science spectra. These fits were computed using version 1.5.9  of \textsc{molecfit} \citep{smette-15}. For this method, the EMIR data format was adapted to \textsc{molecfit} using custom \textsc{python} 3.7 routines, and we masked out from the fit regions with spectral features prominent in K- and M-type stars. We found agreement between the two methods at the 1$\sigma$ level for all the parameters measured in this work. In the paper, we present the results obtained with the \textsc{molecfit} correction, avoiding in this way the use of some low signal-to-noise ratio (SNR) A0~V spectra.

To perform the analysis described in the next sections we imported the telluric-corrected spectra into \textsc{molly}\footnote{\url{http://deneb.astro.warwick.ac.uk/phsaap/software/molly/html/INDEX.html}}. We corrected for the Earth motion and shifted the spectra to the heliocentric rest frame. The spectra were normalised to a spline-fitted continuum and rebinned into a logarithmic scale which provides an uniform velocity scale. The mid-exposure times for each spectrum were calculated and expressed in heliocentric Julian days (UTC).  This was also done for the high SNR ($\geq 80$) library spectra presented in Section~\ref{sec-spectralclassification}. We used them for the spectral classification of the donor star because not all of the observed templates have enough SNR for this task.

\section{Analysis and results}
The measurements in this section were obtained in the wavelength ranges $2.09 - 2.15~$ and $2.18 - 2.27$~$\mu \mathrm{m}$ unless otherwise stated. We did not use the molecular bands at wavelengths $>2.27$~$\mu \mathrm{m}$ due to a much lower SNR. All uncertainties presented in this paper are quoted at $68$ per cent confidence level. Finally, the spectral types assigned to the templates are based on the effective temperature - spectral class correspondence in \cite{pecaut-13}.

\subsection{Radial velocity curve of the donor star}\label{sec-k2}

\begin{figure*}
\centering \includegraphics[height=10cm]{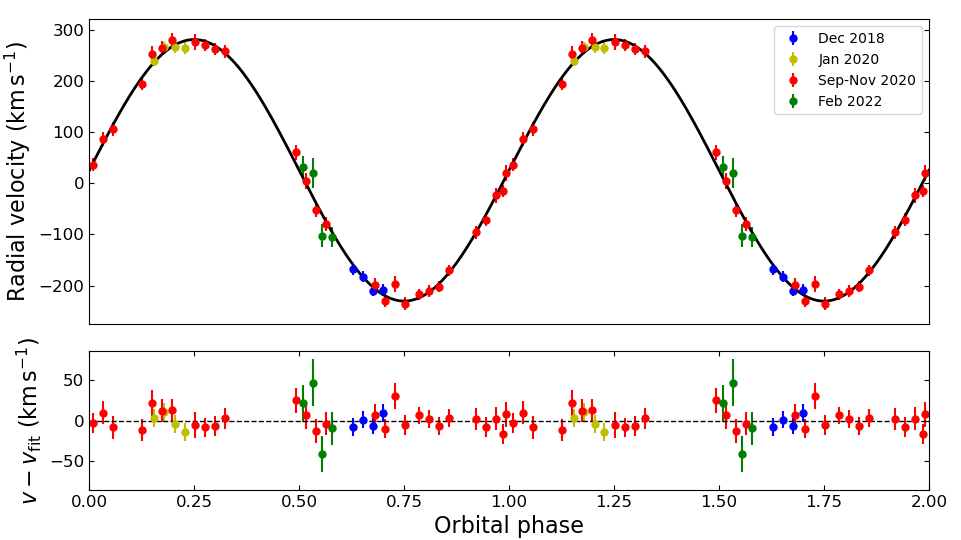}
\caption{\label{fig:radialvelocitycurve} Top panel: Phase-folded heliocentric radial velocity curve of the donor star in \targ\ obtained by cross-correlating the individual spectra with the template spectrum HIP~40910. The best sinusoidal fit is shown as a black line and the orbital cycle has been repeated for the sake of clarity. Bottom panel: residuals of the fit.}
\end{figure*}

We measured the radial velocities by cross-correlating each XY~Ari
spectrum with either the HIP~40375, HIP~40910, HD~157881 or HD~232979 observed templates, that match the spectral type range of the donor star in XY~Ari (more details in Sections~\ref{sec-spectralclassification} and~\ref{sec-rotationalbroadening}). Previous to cross-correlation, the template spectra were corrected for their radial velocities\footnote{HIP~40375 and HD~157881 have \textit{Gaia} radial velocities of $21.393 \pm 0.002$~\kms\ and $-23.523 \pm 0.002$~\kms , respectively \citep{gairv-2018}. According to \cite{chubak-11}, HIP~40910 has a radial velocity of $7.0 \pm 0.2$~\kms . From \cite{spirou-input}, HD~232979 has a radial velocity of $34.2 \pm 0.3$~\kms .}, broadened to match the projected rotational velocity ($v_\mathrm{rot} \sin i$) of the donor star (measured in Section~\ref{sec-rotationalbroadening}) and rebinned onto the same logarithm scale as the XY Ari spectra. We added quadratically the 5~\kms\ uncertainty in the wavelength calibration (Section~\ref{sec-obs}) to the statistical error of each radial velocity measurement. We performed least-squares sinusoidal fits to the radial velocity curves of \targ , $V (t)$, of the form:
\begin{equation}
\label{eq:seno}
V (t) = \gamma + K_{2} \, \mathrm{sin} \, \left[ \frac{2 \pi}{P}(t-T_{0}) \right] ,
\end{equation}  

\noindent where $\gamma$ is  the  heliocentric systemic velocity, $K_2$ the  radial  velocity amplitude of the donor star, $P$ the orbital period and $T_0$ the time of closest approach of the donor star to the observer. The orbital period is determined at $P = 0.25269664(6)$~d from eclipse timing in X-ray and NIR data taken between 1989 and 1995 \citep{allan-96}. Hence, we fixed it in the fits. In Table~\ref{tab:rvcurve_params} we present the best-fit parameters. Note that excluding from the fits the low SNR data from 2022 February does not change the resulting parameters. There are no significant differences between the results using different templates, and the $\chi^{2}$ of the fits is $37-41$ for $37$ degrees of freedom ($\mathrm{dof}$). Hence, we adopt the mean for all the parameters, taking the conservative approach of maintaining the statistical uncertainty from the individual values instead of the standard deviation. Our average values are $\gamma=22 \pm 5$~\kms , $K_2 = 256 \pm 2$~\kms\ and $T_0 \, \mathrm{(HJD)} = 2459122.7116 \pm 0.0005$. Fig.~\ref{fig:radialvelocitycurve} shows the phase-folded radial velocity curve obtained with the HIP~40910 template together with the best-fit sine wave.

Irradiation of the donor star by the WD may affect the $K_2$ measurement by shifting the light centre of the absorption lines (\citealt{hessman-84, marsh-88, wade-88}). To check if this is the case in XY~Ari, we fitted our radial velocity curves using an elliptical orbit. Following \cite{martin-1987}, we used the Wilsing-Russell method \citep[see][]{luyten-1936} suitable for eccentricities $e \leq 0.1$. We find $e = 0.004 \pm 0.012$, which is compatible with null eccentricity and is significantly lower than what is found for CVs with irradiated donor stars like AM Her ($e = 0.05 - 0.07$, \citealt{southwell-95}), IP Peg ($e = 0.089 \pm 0.020$, \citealt{martin-89}) or DO Dra/YY Dra ($e = 0.056 \pm 0.026$, \citealt{mateo-91}). We also fitted a circular orbit to the radial velocities in the orbital phase range $0.8 - 1.2$, when we see the side of the donor that is not facing the WD. This allows to derive $K_2$ when the donor star is irradiated \citep{davey-92, billington-1996}. We obtained $K_2 = 261 \pm 4$~\kms , which agrees at the $1 \sigma$ level with the value obtained by fitting the whole orbit. Based on these two tests, we conclude that our radial velocity curve is unlikely to be affected by heating.

When examining the implications of our $T_0$ value, we found a phase shift of $\simeq 0.11$ cycle relative to the orbital phases predicted by the linear ephemeris of \cite{allan-96} for our data. In this regard, \cite{hellier-97} found an accumulated shift of $0.0077$ when applying the ephemeris to \textit{RXTE} data obtained in 1996, while \cite{zengin-2018} reported a $0.067$ mid-eclipse phase shift when phase-folding \textit{XMM-Newton} data taken in 2010. The uncertainty in the orbital period given by \cite{allan-96} amounts to an accumulated error of $\simeq 0.013$ cycle after $44998$ orbits between their $T_0$ and ours. Possible explanations for this discrepancy are changes in the orbital period over time or an underestimate of the uncertainty due to unaccounted for systematics in Allan et al.'s eclipse time measurements. For instance, the presence of flickering in their NIR light curves could have skewed some of the mid-eclipse time determinations. In addition, their mid-ingress/egress time measurements from X-ray data could be affected by inaccuracies due to not having enough time resolution (see e.g. figure~3 in \citealt{kamata-91}) and/or changes in the relative position of the two X-ray emitting poles between eclipses due to the WD rotation \citep{hellier-97}. To obtain an accumulated error that matches the observed phase shift, the uncertainty in the orbital period should be $\simeq 10$ times higher than reported by \cite{allan-96}, i.e. $10 \times (6 \times 10^{-8} \, \mathrm{d}) \simeq 0.05 \, \mathrm{s}$.

\begin{table}
\caption[]{Donor star radial velocity curve best-fit parameters. The numbers in brackets indicate the uncertainty on the last digit. The 5~\kms\ uncertainty in the wavelength calibration was quadratically added to the uncertainties of the $\gamma$ values. Degrees of freedom ($\mathrm{dof}) = 37$.}
\centering
\begin{tabular}{lccccc}
\hline\noalign{\smallskip}
Template & $\gamma$ & $K_2$ & $T_0$  & $\chi^{2}$/\\
& (\kms) & (\kms) &  (HJD) & $\mathrm{dof}$\\
\hline\noalign{\smallskip}
HIP 40375 & $20(5)$ & $257(2)$ & $2459122.7117(5)$ & 1.03\\
HIP 40910 & $25(5)$ & $256(2)$ & $2459122.7114(5)$ & 0.99\\
HD 157881 & $25(5)$ & $257(2)$ & $2459122.7117(4)$ & 1.11\\
HD 232979 & $16(5)$ & $256(2)$ & $2459122.7115(5)$ & 1.08\\
\hline\noalign{\smallskip}
\end{tabular}
\label{tab:rvcurve_params}
\end{table}

\subsection{Spectral classification of the donor star}\label{sec-spectralclassification}

In order to constrain the spectral type of the donor star in \targ\ we used two grids of high SNR public spectra of K- and M-type stars. The first grid contains thirteen templates with $R = \lambda / \Delta \lambda \simeq 2000$ covering $0.8 - 2.5~\mu \mathrm{m}$ and extracted from the InfraRed Telescope Facility (IRTF) spectral library \citep{rayner-2009}. For K-type dwarfs we took the average photometric effective temperature estimates provided in \cite{luck-2017}, while for M-dwarfs we used the photometric effective temperatures given in \cite{houdebine-19}. This was not possible for two objects (HD~45977 and HD~237903), for which we adopted the temperature values from \cite{sousa-2011} and \cite{niedzielski-2016}, respectively. The resolution of this grid is comparable to the broadening of
the XY Ari data resulting from convolution of the instrumental and the rotational profiles. The second grid includes eight high resolution ($R = \lambda / \Delta \lambda \simeq 10000$) spectra from the X-shooter Spectral Library (XSL) Data Release 2 \citep{gonneau-2020}. The effective temperatures were spectroscopically derived by \cite{arentsen-19}. From visual inspection, we selected the XSL spectra less affected by emmision spike artifacts in the wavelength regions used in our analysis. However, all of them suffer from artifacts at $\simeq 2.26-2.28~\mu \mathrm{m}$ due to imperfect order merging. Thus, we were forced to perform the analysis in the $2.09 - 2.15~$ and $2.18 - 2.25$~$\mu \mathrm{m}$ regions with the caveat that compared to the IRTF templates we do not cover the Ca~\textsc{i} photospheric lines. Table~\ref{tab:irtf_xsl} lists all the IRTF and XSL selected templates.

We applied the optimal subtraction technique described in \cite{marsh-94} to all the \targ\ spectra with both grids of templates. It searches for the smoothest residual obtained after subtracting a set of normalized spectral templates from the average, Doppler-corrected and normalized spectrum of the target (see e.g. Fig.~\ref{fig:optsub}). In the case of the XSL grid, we previously downgraded the spectral templates to match the resolution of the \targ\ spectra by convolution with a Gaussian profile with $\mathrm{FWHM}=63$~\kms . We summarize here how we applied the optimal subtraction technique with each of the templates. First, we velocity-shifted each \targ\ spectrum to the rest frame of the template. Next, we computed a weighted average of the \targ\ spectra giving larger weights to those with higher SNR. We subsequently broadened the photospheric lines of the template spectrum by convolution with the Gray's rotational profile \citep{libro-gray}. We adopt a linear limb-darkening coefficient for the $K$-band of $0.25$, a reasonable choice for a late K-type star \citep{claret95} and probed the $v_\mathrm{rot} \sin i$ space between $1$ and $200$\,\kms\ in steps  of $1$\,\kms. A robust measurement of $v_\mathrm{rot} \sin i$ will be given in Section~\ref{sec-rotationalbroadening}, where we will use templates taken with the same instrumental setup as the XY~Ari data. The broadened versions of each template were multiplied by a factor $f$ between $0$ and $1$ and then subtracted from the weighted-average spectrum of \targ. This factor represents the fractional contribution of the donor star to the total flux in the wavelength range of the analysis. Finally, we searched for the values of $v_\mathrm{rot} \sin  i$ and $f$ that minimised the $\chi^2$ between the residual of the subtraction and a smoothed version of itself obtained by convolution with a Gaussian profile with $\mathrm{FWHM}=30$~\AA . This smoothing removes large-scale trends in the residual spectrum.

\begin{figure}
\centering \includegraphics[height=9.5cm]{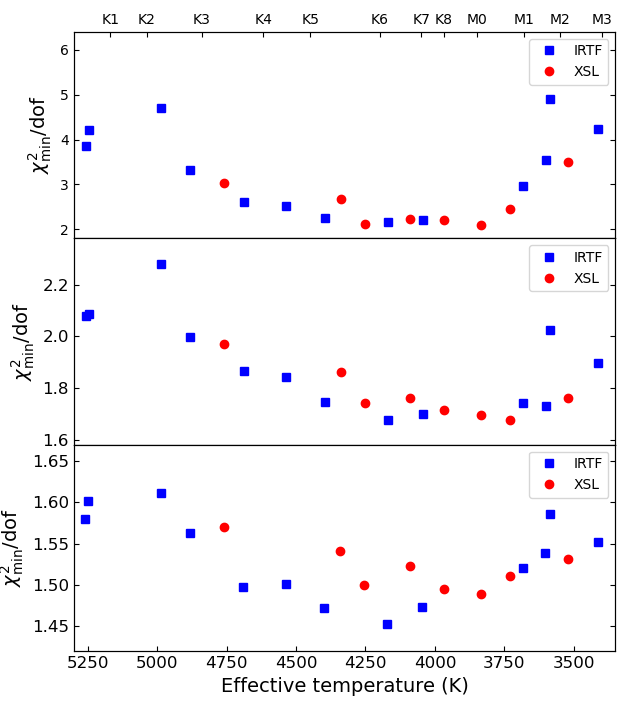}
\caption{\label{fig:spc-class} Spectral classification of the donor star using the optimal subtraction technique with template spectra from IRTF (blue squares) and XSL (red circles). Results are shown for the full average spectrum of XY Ari (top panel) and averages at
orbital phases $0.9-0.1$ (middle panel) and $0.4-0.6$ (bottom panel). The spectral type -- effective temperature correspondence is indicated.}
\end{figure}

\begin{figure*}
\centering \includegraphics[height=10.5cm]{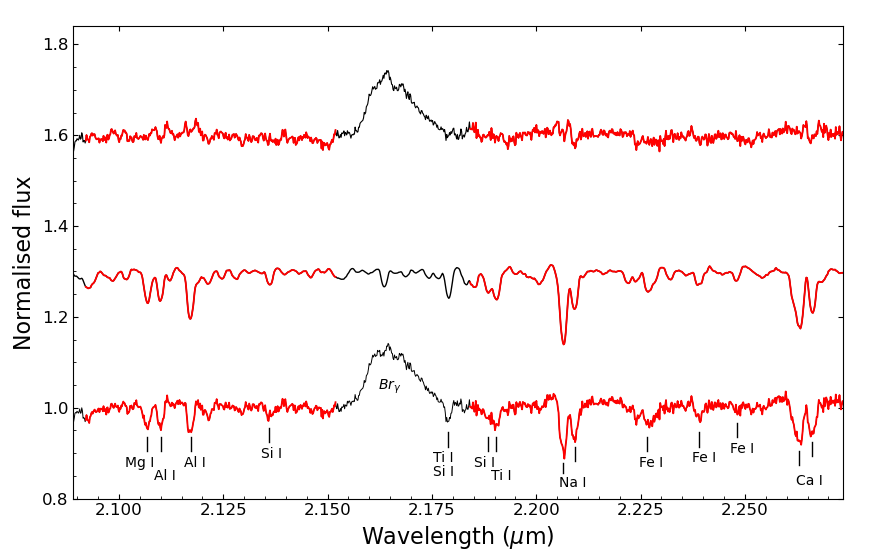}
\caption{\label{fig:optsub} Illustration of the optimal subtraction technique. From bottom to top: \targ\ average spectrum in the rest frame of the donor star, spectrum of the HD~157881 template (K6--K7~V) observed with the same instrument setup and broadened to match the donor star's $v_\mathrm{rot} \sin  i$, and the residual after subtraction of the broadened and scaled template. The wavelength regions used for the analysis presented in Sections~\ref{sec-k2}, ~\ref{sec-spectralclassification} (for the IRTF templates) and ~\ref{sec-rotationalbroadening} are marked in red. The template and residual spectra have been shifted vertically for display purposes. The main spectral features are identified according to the NASA IRTF spectral library atlas \citep{rayner-2009}.}
\end{figure*}

The top panel in Fig.~\ref{fig:spc-class} shows the minimum $\chi^{2}/\mathrm{dof}$ for each template. The minimization of $\chi^{2}/\mathrm{dof}$ shows a plateau in the effective temperature range $3850-4250$~K. Visual inspection of the broadened versions of the IRTF/XSL templates in that range and the XY~Ari average spectrum did not allow to obtain a narrower constraint for the spectral type of the donor. According to \cite{pecaut-13}, the $3850-4250$~K effective temperature range corresponds to K6 -- M0~V. \cite{littlefair-2001} applied a similar $\chi^{2}$-based analysis employing the average of $K$-band spectra obtained with a $\simeq 350$~\kms\ resolution that covered orbital phases $0.64-0.18$. They found a K7 -- M0~V donor star, which agrees with our constraint.

We searched for differences in the temperature of the donor's day and night hemispheres. To do so, we produced average spectra at orbital phases $0.9-0.1$ (night side, from eight individual spectra) and $0.4-0.6$ (day side, from eight individual spectra, being four of them those with low SNR taken on 2022 February). We then repeated the optimal subtraction process for both cases. The middle and bottom panels of Fig.~\ref{fig:spc-class} show the results for phases $0.9-0.1$ and $0.4-0.6$, respectively. Compared to the results of the full average, the $\chi^{2}_{\mathrm{min}}/\mathrm{dof}$--effective temperature relation is less abrupt for the day- and night-side averages as a consequence of the lower SNR. Nevertheless, the IRTF grid template spectra (covering more photospheric lines than the XSL grid) show that the lowest $\chi^{2}_{\mathrm{min}}/\mathrm{dof}$ corresponds to the template HD~201092 for both the day- and night-side averages. This template has an effective temperature of $4171 \pm 39$~K,  in line with a K6~V spectral type. We consider this, together with the null eccentricity in the radial velocity curve (Section~\ref{sec-k2}), as evidence of no significant heating of the donor star photosphere. \cite{allan-96} reached the same conclusion based on the study of their $H$-band ellipsoidal+eclipse light curve. As a final test, we also looked for a narrow component in the Br$\gamma$ emission line, that may originate on the irradiated face of the donor star if heating were significant. For this we inspected the emission-line trailed spectrogram (not shown), but we did not detect any sign of such a feature.

\subsection{Projected rotational velocity of the donor star}\label{sec-rotationalbroadening}

\begin{table}
\caption[]{IRTF and XSL templates used for the spectral classification.}
\centering
\begin{tabular}{lcccccc}
\hline\noalign{\smallskip}
Template & Library & Sp. & Effective & Ref. \\
\tabhead{} & & Type & temperature & \\
\tabhead{} &  &  & (K) & \\
\hline\noalign{\smallskip}
HD 10476 & IRTF & K0 V  & $5259 \pm 55$ & $a$ \\
HD 145675 & IRTF & K0 V & $5248 \pm 79$ & $a$ \\
HD 3765 & IRTF & K2.5 V & $4987 \pm 32$ & $a$ \\
HD 219134 & IRTF & K3 V & $4883 \pm 56$ & $a$ \\
HD 45977 & IRTF & K4 V & $4689 \pm 174$ & $b$ \\
HD 36003 & IRTF & K4.5 V & $4536 \pm 28$ & $a$ \\
HD 237903 & IRTF & K5 V & $4398 \pm 32$ & $c$ \\
HD 201092 & IRTF & K6 V & $4171 \pm 39$ & $a$ \\
HD 19305  & IRTF & K7 V & $4045 \pm 13$ & $d$ \\
HD 42581  & IRTF & M1 V & $3683 \pm 13$ & $d$ \\
HD 95735  & IRTF & M1.5 V & $3603 \pm 13$ & $d$ \\
Gl 806  & IRTF & M2 V & $3587 \pm 13$ & $d$ \\
Gl 388  & IRTF & M3 V & $3414 \pm 13$ & $d$ \\
HD 218566  & XSL & K3.5 V & $4761 \pm 50$ & $e$ \\
HD 21197  & XSL & K5.5 V & $4340 \pm 45$ & $e$ \\
HD 118100  & XSL & K6 V & $4255 \pm 44$ & $e$ \\
HD 204587  & XSL & K7 V & $4091 \pm 43$ & $e$ \\
HIP 70472  & XSL & K8 V & $3968 \pm 52$ & $e$ \\
HIP 100047  & XSL & M0 V & $3835 \pm 50$ & $e$ \\
HD 209290  & XSL & M0.5 V & $3731 \pm 59$ & $e$ \\
HD 119850  & XSL & M2 V & $3521 \pm 56$ & $e$ \\
\hline\noalign{\smallskip}
\multicolumn{2}{l}{\footnotesize{$^a$\cite{luck-2015}}} \\
\multicolumn{2}{l}{\footnotesize{$^b$\cite{sousa-2011}}} \\
\multicolumn{2}{l}{\footnotesize{$^c$\cite{niedzielski-2016}}} \\
\multicolumn{2}{l}{\footnotesize{$^d$\cite{houdebine-19}}} \\
\multicolumn{2}{l}{\footnotesize{$^e$\cite{arentsen-19}}} \\
\end{tabular}
\label{tab:irtf_xsl}
\end{table}

\begin{table}
\caption[]{$v_\mathrm{rot} \sin  i$ and $f$ from optimal subtraction of the template spectra observed with the same instrumental setup employed for XY~Ari. The numbers in brackets indicate the uncertainties on the last digit.}
\centering
\begin{tabular}{lcccccc}
\hline\noalign{\smallskip}
Template & Sp. & Effective & $v_\mathrm{rot} \sin  i$ & $f$  \\
\tabhead{} & Type & temperature & (\kms) & \\
\tabhead{} &  & (K) &  \\
\hline\noalign{\smallskip}
HIP 40375 & K4--K7 V & $4334 \pm 209$  & $142(3)$ & $0.78(2)$ \\
HIP 40910 & K5--K8 V & $4159 \pm 228$  & $142(3)$ & $0.83(2)$  \\
HD 157881 & K6--K7 V & $4124 \pm 60$ & $139(3)$ & $0.65(2)$ \\
HD 232979 & K8--K9 V & $3929 \pm 60$ & $141(4)$ & $0.72(2)$ \\
\hline\noalign{\smallskip}
\end{tabular}
\label{tab:spectraltype}
\end{table}

We exploit again the optimal subtraction technique, this time to measure $v_\mathrm{rot} \sin  i$ using four of the spectral templates of dwarfs taken with EMIR (Section~\ref{sec-obs}). These templates, listed in Table~\ref{tab:spectraltype}, are affected by the same instrumental broadening as the target spectra, making them ideal to properly measure $v_\mathrm{rot} \sin  i$. The effective temperature of these templates is within the constraint for the donor star of XY~Ari found in the previous section.

For the measurement of $v_\mathrm{rot} \sin  i$ we simulated the effect of the drift in the donor's absorption lines during the exposures by making one copy of a given template for each science spectrum, smearing them according to the effective length of the science exposures and the orbital parameters, and finally computing an average from those copies (see e.g. \citealt{torres-2002}). 

We compared the results obtained using different FWHM values (between $1$ and $50$~\AA) of the smoothing Gaussian that is applied to the residual during the optimal subtraction process. The impact of the FWHM on $v_\mathrm{rot} \sin  i$ was $<5$~\kms\ with a weak parabolic relationship between the resulting $v_\mathrm{rot} \sin  i$ and the selected FWHM, with a minimum at $\simeq 30$~\AA. Performing this analysis with four artificially broadened templates resulted in very similar trends and showed that the minimum $v_\mathrm{rot} \sin  i$ is the actual broadening applied to the data. Hence, we finally selected the results obtained using a 30-\AA \, FWHM smoothing Gaussian.

Evaluation of the uncertainties in $v_\mathrm{rot} \sin  i$ and $f$ was performed by Monte Carlo randomization following the approach in \cite{steeghs-2007} and \cite{torres-2020}. For each template, the optimal subtraction procedure was repeated for 10000 bootstrapped copies of the \targ\ average spectrum. This delivered normal distributions of possible values for $v_\mathrm{rot} \sin  i$ and $f$, so we took their mean and standard deviation as the value and $1 \sigma$ uncertainty, respectively.

The best $v_\mathrm{rot} \sin  i$ and $f$ for each template are presented in Table~\ref{tab:spectraltype}. Based on these measurements, we adopt a mean $v_\mathrm{rot} \sin  i = 141 \pm 3$~\kms\ , deriving the uncertainty as we did for the parameters presented in Section~\ref{sec-k2}. In this case, this conservative approach serves to account for the potential systematic uncertainties induced by the choice of the limb-darkening coefficient. In this regard, using values of $0$ and $0.5$ changes the $v_\mathrm{rot} \sin  i$ by $-3$ and $+3$~\kms , respectively. Our \targ\ spectra do not cover the full orbital cycle, but they have a pretty even sampling of the orbital phases critical to account for the $v_\mathrm{rot} \sin  i$ variability. Thus, the $v_\mathrm{rot} \sin  i$ value presented here must serve as a good measure of its mean value around the full orbit. In fact, we arrive at a similar result adopting the mean value of the $v_\mathrm{rot} \sin  i$ measurements obtained from the spectra observed at the $0.9-0.1$, $0.15-0.35$, $0.4-0.6$ and $0.65-0.85$ orbital phase ranges. From the measurements presented in Table~\ref{tab:spectraltype} we also conclude that the donor star contributes $\simeq70-80$ per cent to the total flux in the $K_\mathrm{s}$-band at the time of our spectroscopy.

\subsection{Binary mass ratio, orbital inclination and stellar masses}\label{sec-masses}

We derived the donor-to-WD mass ratio ($q=M_2/M_1$) for \targ\ using the equation:
\begin{equation}
\label{eq:vsini}
v_\mathrm{rot} \sin  i  \simeq 0.49 (1+q) q^{2/3}  K_2 \left[0.6 q^{2/3} + \mathrm{ln} (1+q^{1/3})  \right]^{-1} \, ,
\end{equation}

\noindent which is obtained using a spherical aproximation for the Roche Lobe radius of the donor star \citep{eggleton-83} and making some natural assumptions for the system: circular orbit, synchronized rotation of the donor star with the orbital motion, and alignment of the orbital and stellar spin axes. To compute the uncertainty in $q$ we followed a Monte Carlo approach that picked random values of $K_2$ and $v_\mathrm{rot} \sin  i$ from normal distributions defined by the mean and the $1 \sigma$ uncertainties of our measurements. We then constructed a probability distribution for $q$ by calculating its value for 10000 random sets of $K_2$ and $v_\mathrm{rot} \sin  i$. This distribution was well fitted with a Gaussian, so we took its mean and standard deviation as reliable estimates of $q$ and its uncertainty, respectively. We derive below the uncertainties in $i$, $M_1$ and $M_2$ following the same Monte Carlo approach.

Using the values of $K_2$ and $v_\mathrm{rot} \sin  i$ derived in Sections~\ref{sec-k2} and~\ref{sec-rotationalbroadening} in Eq.~\ref{eq:vsini}, we obtain 
\begin{equation}
\label{eq:q_resultado}
q = 0.62 \pm 0.02, 
\end{equation}

\noindent which is within the $0.48 < q < 0.68$ range proposed by \cite{hellier-97}.

The orbital inclination, $i$, can be inferred from $q$ and the width of the X-ray eclipse considering when a  point is eclipsed by the photosphere of the
Roche-lobe-filling donor star. \cite{chanan-76} set
through ray tracing the eclipse limit conditions in their eq.~4 and 5. We solve this system of non-linear equations following the
minimization approach described by the authors. Using the derived
$q$, an average mid-ingress to mid-egress eclipse duration of $2050 \pm 28$~s \citep{kamata-91,allan-96} and the Broyden–Fletcher–Goldfarb–Shanno algorithm
we obtain:\footnote{A fully consistent inclination is derived from the computation of a synthetic light curve considering the Roche lobe geometry and matching the eclipse duration. On the other hand, its analytical calculation under the approximation of an spherical donor with an effective Roche lobe
radius (e.g. eq.~2 in \citealt{dhillon-92}) gives $i = 79.6^{\circ} \pm 0.5^{\circ}$, that underestimates our numerically calculated value by $1.2^{\circ}$.}
\begin{equation}
\label{eq:i_resultado}
i = 80.8^{\circ} \pm 0.5^{\circ} ,
\end{equation}

\noindent which agrees with previous constraints of $80^{\circ} < i < 87^{\circ}$ \citep{allan-96} and $80^{\circ} < i < 84^{\circ}$ \citep{hellier-97}.

Finally, from our $K_2$, $q$ and $i$ values and the orbital period of \cite{allan-96}, considering a $\pm 0.05$~s uncertainty, we obtain the masses of the stars in \targ\ to be:
\begin{equation}
\label{eq:m1}
M_{1}=\frac{P (1+q)^{2}}{2\pi \mathrm{G}}\frac{K_{2}^{3}}{\sin^{3}i} = 1.21 \pm 0.04 \, \mathrm{M}_{\odot} 
\end{equation}  

\begin{equation}
\label{eq:m2}
M_{2} = q M_{1} = 0.75 \pm 0.04 \, \mathrm{M}_{\odot} .
\end{equation}  

\noindent The inclination was derived under the assumption that the source of the X-ray emission eclipsed by the donor star is centred on the WD. This is unlikely, since the WD in XY Ari shows during quiescence two nearly opposite X-ray emitting poles. They are at a latitude between $\simeq 44^{\circ}$ and $\simeq 63^{\circ}$, tracing the asynchronous WD rotation \citep{hellier-97}. Consequently, the eclipse duration can vary depending on the relative position of the poles on the WD surface. For instance, at the start of a given eclipse, one pole could appear over the WD limb, while at the start of another eclipse it could be above the WD centre. In the first example, the eclipse would start sooner, producing a longer eclipse. However, the eclipse length variations are limited, considering that it takes the donor star between $22$ and $35$~s to travel the entire WD diameter \citep{hellier-97}, and the change in the relative position of the poles is a fraction of that distance. Furthermore, the eclipse duration reported by \cite{allan-96} is the mean value obtained from the observation of six X-ray eclipses. While it would be ideal to obtain the duration from a very large number of X-ray eclipse observations with higher time resolution, we consider that the eclipse length in \cite{allan-96} is accurate enough for our purposes. In fact, even if it were biased by $\pm 50$~s, which is comparable to the length of the X-ray eclipse ingress/egress \citep{kamata-91, hellier-97, zengin-2018}, we would obtain an inclination $i = 80.2^{\circ} \pm 0.5^{\circ}$ for a $2000 \pm 28$~s eclipse duration and $i = 81.6^{\circ} \pm 0.5^{\circ}$ for $2100 \pm 28$~s. This change in the inclination would have a $<1$~per cent effect on the dynamical masses, which illustrates the robustness of our results.

\comment{
\textbf{On the other hand, the $2050 \pm 28$~s mid-ingress to mid-egress eclipse duration reported by \cite{allan-96} implies that the ingress/egress takes $60$~s. Other authors have measured the ingress/egress from X-ray light curves with higher time resolution and have found lower values (e.g. .....). comes from the average of X eclipses. Therefore, the statistical error includes this effect. Moreover, the ingress and egress takes $\simeq 60$ according to \cite{allan-96}.   Note that, given the derived high inclination, even if the uncertainty in $i$ were $2^{\circ}$, it would only translate into a $14$ and $4$ per cent increment in the uncertainties of $M_{1}$ and $M_{2}$, respectively.}}

\section{Discussion}\label{sec-discuss}

\subsection{Stellar masses}
The donor in \targ\ is consistent with a spectral type K6--M0~V. It has a mass $M_{2}= 0.75 \pm 0.04 \, \mathrm{M}_{\odot}$ and its Roche lobe volume radius can be calculated as $R_2 = \frac{P}{2 \pi} \frac{v_\mathrm{rot} \sin  i} {\mathrm{sin} \, i} = 0.71 \pm 0.02 \, R_{\odot}$. The mass and radius are too high for a typical M0~V star, but they agree with the canonical mass and radius of a K6~V star, which are $\simeq 0.7 \, \mathrm{M}_{\odot}$ and $\simeq 0.7 \, \mathrm{R}_{\odot}$, respectively \citep{pecaut-13}. We also derive a surface gravity $\log \, g = 4.6 \pm 0.03$~dex. Based on these parameters, the donor star better fits a main sequence K6 star. Note that when we performed the spectral classification of the donor star (Section~\ref{sec-spectralclassification}) we found that the best-matching IRTF template has an effective temperature of $4171 \pm 39$~K, which corresponds to a K6~V star according to \cite{pecaut-13}.

\cite{zorotovic-2011} compiled all the WD masses from the literature that they considered as robust determinations\footnote{The methods considered as reliable by these authors are: analysis of eclipse light curves / radial velocity curves, derivation of the gravitational redshift from the systemic velocity of both the WD and the donor star, and modelling of the WD ultraviolet spectrum.}. Their sample contains 32~CVs with an average WD mass of $0.82 \pm 0.15 \, \mathrm{M}_{\odot}$. \cite{pala-2022} considered the same criterion and updated the census to 54~CVs (see table~44 in the online supplementary material of their work). They also used high quality \textit{Hubble Space Telescope} ultraviolet spectra and \textit{Gaia} parallaxes to extend the sample to 89 systems, obtaining an average WD mass of $0.81^{+0.16}_{-0.20} \, \mathrm{M}_{\odot}$. The WD in \targ\ is therefore more massive than the average at $> 2 \sigma$. In fact, it would be among the three most massive WDs if included in the total sample of 89 CVs (now 90) from \cite{pala-2022}. The other two systems are the dwarf novae U~Gem ($M_1 = 1.20 \pm 0.05 \, \mathrm{M}_{\odot}$) and IP Peg ($M_1 = 1.16 \pm 0.02 \, \mathrm{M}_{\odot}$).

\begin{figure}
\centering \includegraphics[height=10cm]{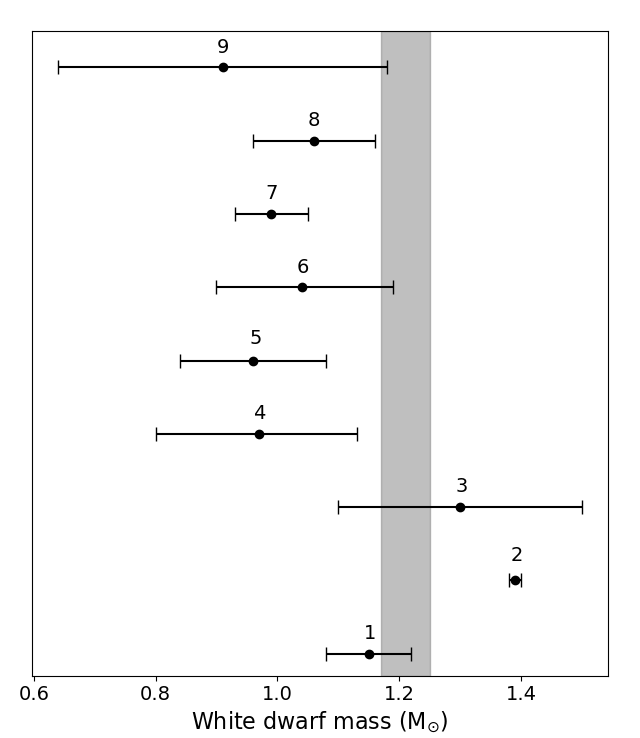}
\caption{\label{fig:masses} Comparison between the dynamical mass of the WD in \targ\ presented in this paper (shaded area, $1 \sigma$) and previous estimates obtained using spectral modelling of: \textit{RXTE}/\textit{ASCA}/\textit{GINGA} soft X-ray data (labelled in the figure as $1$, $2$, and $3$, respectively; \citealt{ramsay-98}), \textit{RXTE} soft X-ray data ($4$; \citealt{ramsay-2000}), \textit{Swift}  hard X-ray data ($5$ and $8$; \citealt{brunchsweiger-2009} and \citealt{suleimanov-19}), \textit{Suzaku} hard X-ray data ($6$; \citealt{yuasa-2010}) and \textit{NuSTAR} hard X-ray data ($7$; \citealt{xu-2019}). An estimate based on the ratio of the iron emission lines in \textit{NuSTAR} and \textit{Suzaku} spectra is also included ($9$; \citealt{xu-2019}).}
\end{figure}

\subsection{An updated estimate of the distance to XY~Ari}\label{subsec-distance}

Here we will exploit the constraint on the donor  spectral classification and the photometry from the EMIR acquisition images to improve the distance estimate
presented by \cite{littlefair-2001}. From the multi-epoch target acquisitions we derive a phase-average magnitude of $K_{\mathrm{s}}=13.4 \pm 0.2$, while from the single acquisition done in the $J$-band we obtain $J=16.4 \pm 0.2$. Using the ratio between the extinction in a given band and the colour excess, $A_{\lambda} / E(B-V)$, from \citep{schlafly-2011}, we derive:
\begin{equation}
\label{eq:color_excess}
m_{K_{\mathrm{s}}}^{\mathrm{o}} - m_{J}^{\mathrm{o}} = (M_{K_{\mathrm{s}}} - M_J) - 0.407 \times E(B-V) ,
\end{equation} 

\noindent where $m_{\lambda}^{\mathrm{o}}$ and $M_{\lambda}$ are the observed (reddened) apparent and absolute magnitudes of the donor star, respectively. Following \cite{littlefair-2001}, we assume  equal fractional light contribution $f$ of the donor star to the total flux in the $K_{\mathrm{s}}$- and $J$-bands. Thus, using a flat probability distribution for the spectroscopic constraint on $f$ ($=0.7-0.8$; Section~\ref{sec-rotationalbroadening}) we derive $m_{K_{\mathrm{s}}}^{\mathrm{o}} = 13.7 \pm 0.2$ and $m_{J}^{\mathrm{o}} = 16.7 \pm 0.2$. Adopting a K6~V donor star and employing the corresponding absolute magnitudes ($M_{K_{\mathrm{s}}} = 4.56$, $M_J = 5.31$; \citealt{pecaut-13}), we obtain a colour excess $E(B-V) = 5.5 \pm 0.7$. From the colour excess we calculate the extinction in the $K_{\mathrm{s}}$-band: $A_{K_{\mathrm{s}}} = 0.302 \times E(B-V) = 1.7 \pm 0.2$ \citep{schlafly-2011}. The intrinsic apparent magnitude of the likely K6~V donor star in XY Ari is $m^{\mathrm{i}}_{K_{\mathrm{s}}} = m_{K_{\mathrm{s}}}^{\mathrm{o}} - A_{K_{\mathrm{s}}} = 12.0 \pm 0.3$. Using $m^{\mathrm{i}}_{K_{\mathrm{s}}}$ and $M_{K_{\mathrm{s}}}$ in the distance modulus gives $d = 308 \pm 40$~pc. In comparison, a distance of $270 \pm 100$~pc was found by \cite{littlefair-2001} based on less-constrained parameters and characterization of the donor star.

\subsection{Comparison with the WD mass from X-ray spectral modelling}\label{subsec-m1x}

In IPs, the temperature of the plasma in the region where the accreted material impacts the WD surface is thought to depend mainly on the WD mass and radius \citep{aizu-73}, which are linked through the theoretical mass-radius relation for WDs \citep{hamada-61,Nauenberg-72,panei-2000}. The main cooling mechanism of the accreting region is assumed to be the emission of X-rays and, therefore, modelling of the X-ray spectrum can deliver an estimate of its WD mass (hereinafter $M_1^\mathrm{X}$). Some authors have proposed that comptonization could be the main mechanism behind the X-ray emission in IPs (e.g. \citealt{maiolino-21}), but the most extended and accepted models are based on thermal bremsstrahlung.
\comment{Many authors \citep{kamata-93,salinas-2004,norton-2007,zengin-2018} have presented fits to $<20$~keV X-ray spectra of \targ\ with thermal bremsstrahlung models. They found temperatures in the range $kT_\mathrm{s} \simeq 20-40$~keV, while a $1 \, \mathrm{M}_{\odot}$ WD would imply a shock temperature of $\simeq 60$~keV \citep{zengin-2018}. In this regard, \cite{suleimanov-2005} proposed that the use of the X-ray continuum in the energy range $<20$~keV (hereinafter soft X-rays) may not provide an accurate estimate of the WD mass in IPs, since for masses $>0.6 \, \mathrm{M}_{\odot}$ the post-shock temperature is $kT_\mathrm{s}>20$~keV. }

\cite{ramsay-98} performed a multi-instrument X-ray study of \targ\ employing soft X-ray spectra and a multi-temperature bremsstrahlung model. They found instrument depending results in spite of the similar energy ranges. From $2-20$~keV \textit{RXTE/PCA} data they derived $M_1^\mathrm{X} = 1.15 \pm 0.07 \, \mathrm{M}_{\odot}$, while from $1-12$~keV \textit{ASCA} data (two or more detectors) the result was $M_1^\mathrm{X} = 1.39 \pm 0.01 \, \mathrm{M}_{\odot}$. Finally, from $2-20$~keV \textit{Ginga/LAC} data the derived WD mass was $M_1^\mathrm{X} = 1.3 \pm 0.2 \, \mathrm{M}_{\odot}$. These measurements average to $M_1^\mathrm{X} = 1.28 \pm 0.04 \, \mathrm{M}_{\odot}$, which they deemed an overestimate since they expected a WD mass of $1 \, \mathrm{M}_{\odot}$ based on assumptions such as a WD radius of $4.3-7.3 \times 10^8$~cm \citep{hellier-97} and $M_{2}= 0.62 \, \mathrm{M}_{\odot}$ (from Patterson's mass-radius relation, \citealt{patterson-1984}). Our dynamical study reveals that, in fact, their mean $M_1^\mathrm{X}$ agrees within $1 \sigma$ of our WD mass. However, it should be noted that this average X-ray value comes from inconsistent estimates. Later, \cite{ramsay-2000} used a refined version of the previous model to fit $2-20$~keV \textit{RXTE/PCA} spectra of several IPs. For XY~Ari, this improved model yielded $M_1^\mathrm{X} = 0.97^{+0.16}_{-0.17} \, \mathrm{M}_{\odot}$, which only agrees at the $2 \sigma$ level with our dynamical measurement. In general, soft ($<20$~keV) X-ray spectra do not provide accurate estimates for WD masses in IPs, since for masses $>0.6$\,M$_{\odot}$ the maximum post-shock temperature is $>20$~keV \citep{suleimanov-2005}. Further, the soft spectrum of many IPs is strongly affected by absorption and reflection components that are still not well understood \citep{suleimanov-19}.

WD mass estimates for XY~Ari have also been derived from hard ($>20$~keV) X-ray data.  \cite{brunchsweiger-2009} presented a fit to a $15-195$~keV \textit{Swift}/BAT spectrum of \targ\ employing the multi-temperature bremsstrahlung model of \cite{suleimanov-2005} and obtained $M_1^\mathrm{X} = 0.96 \pm 0.12 \, \mathrm{M}_{\odot}$. Likewise, \cite{yuasa-2010} used an improved version of the previous model with $3-50$~keV \textit{Suzaku} data and found $M_1^\mathrm{X} = 1.04^{+0.15}_{-0.14} \, \mathrm{M}_{\odot}$. Following a similar approach, \cite{xu-2019} modelled a $3-50$~keV \textit{NuSTAR} spectrum and derived $M_1^\mathrm{X} = 0.99 \pm 0.06 \, \mathrm{M}_{\odot}$. Finally, \cite{suleimanov-19} employed a more complex model that considers a finite fall height for the accreted matter. From \textit{Swift}/BAT XY~Ari data they obtained $M_1^\mathrm{X} = 1.06 \pm 0.10 \, \mathrm{M}_{\odot}$, which is below the dynamical WD mass, but is very close to a $1 \sigma$ level agreement. These comparisons show that the $M_1^\mathrm{X}$ values derived from hard X-ray spectral bands (which are supposed to be more reliable) may well underestimate the mass of the WD in XY Ari. In this respect, \cite{ayoze-2021} found that most of the estimates with uncertainties $<0.2\, \mathrm{M}_{\odot}$ obtained from X-ray spectral fitting for the WD mass in the IP GK~Per (dynamical mass $M_1 \simeq 1 \, \mathrm{M}_{\odot}$) are underestimates, while different indirect techniques such as deriving the quiescence-to-outburst Alfvén radius ratio from X-ray spectra or modelling the nova event optical light curve, provided accurate values.

Other methods have been explored to derive WD masses in IPs using X-ray data. In particular, \cite{ezuka-99} used a $5-10$~keV \textit{ASCA} spectrum of \targ\ to derive a temperature of $4.9^{+9.9}_{-4.9}$~keV from the intensity ratio of the $6.7$~and $7.0$~keV iron emission lines and theoretical relations from \cite{mewe-85}. They also fitted the continuum, obtaining $16^{+21}_{-6}$~keV. Similar temperature differences appeared for most magnetic CVs in their study, as a consequence of the temperature gradient in the plasma. The continuum temperature was expected to represent an average of the temperature distribution, so they derived a relation between both temperature determinations and used it to derive the WD masses from the iron emission line ratios. Unfortunately, they did not provide a value for \targ . We guess the reasons for that were poorly resolved emission lines and that XY Ari clearly deviates from the relation between temperature determinations (see the position of \targ\ in their fig.~4). More recently, \cite{xu-2019} derived a relation between WD masses in IPs and the intensity ratio of the $6.7$ and $7.0$~keV iron emission lines based on synthetic (model-generated) spectra. They derived $M_1^\mathrm{X} = 0.91 \pm 0.27 \, \mathrm{M}_{\odot}$ for \targ , whose uncertainty is too high to draw any conclusion.

In brief, a number of the $M_1^\mathrm{X}$ values reported are not in good agreement with our dynamical WD mass. Figure~\ref{fig:masses} gives a visual summary of the contents of this section. In view of our results and those presented in \cite{ayoze-2021} for GK~Per, we warn that WD average masses in IPs obtained from X-ray spectral modelling should be taken with caution. The improvement of hard X-ray spectrometers and refined accretion models have allowed to reduce very significantly the statistical uncertainties in the $M_1^\mathrm{X}$ values (e.g. \citealt{shaw-2020}), but there still exist important systematic deviations that must be addressed. In this regard, \cite{belloni-21} suggested that X-ray spectral modelling alone may not be enough to derive reliable physical parameters. They showed that there is a degeneracy in the parameter space of X-ray spectral models. Multiple combinations of the parameters (WD mass, magnetic field strength, magnetospheric radius and specific accretion rate) can produce the same spectrum. According to these authors, additional data, such as simultaneous X-ray light curves, are required to break the degeneracy and obtain accurate estimates of the parameters.

\section{Conclusions}\label{sec-conclusions}

\begin{table}
\caption[]{Fundamental parameters of \targ\ presented in this article.}
\label{tab:parameters_summary}
\centering
\begin{tabular}{l c }
\hline\noalign{\smallskip}
$K_2 \, (\mathrm{km~s^{-1}})$ & $256 \pm 2$ \\
\noalign{\smallskip}
$v_\mathrm{rot} \sin  i \, (\mathrm{km~s^{-1}})$ & $141 \pm 3$ \\
\noalign{\smallskip}
$q$ & $0.62 \pm 0.02$ \\
\noalign{\smallskip}
$i \,(^{\circ})$ & $80.8 \pm 0.5$ \\
\noalign{\smallskip}
$M_1 \, (\mathrm{M}_{\odot})$ & $1.21 \pm 0.04$\\
\noalign{\smallskip}
$M_2 \, (\mathrm{M}_{\odot})$ & $0.75 \pm 0.04$\\
\noalign{\smallskip}
$R_2 \, (\mathrm{R}_{\odot})$ & $0.71 \pm 0.02$\\
$d$ \, (pc) & $308 \pm 40$\\
\hline\noalign{\smallskip}
\end{tabular}
\end{table}

We have performed a dynamical study of the eclipsing IP \targ\ using GTC/EMIR NIR spectra taken in $2018-2022$. We constrain the spectral type of the donor star, derive its radial velocity curve and measure its rotational broadening. We found that the donor star is consistent with a K6~V star with $K_2=256 \pm 2$~\kms\ and $v_\mathrm{rot} \sin  i = 141 \pm 3$~\kms . Using these values we derived the binary mass ratio $q = 0.62 \pm 0.02$, which allowed us to better constrain the orbital inclination, $i = 80.8^{\circ} \pm 0.5^{\circ}$. We finally derive the dynamical masses $M_1 = 1.21 \pm 0.04 \, \mathrm{M}_{\odot}$ for the WD and $M_2 = 0.75 \pm 0.04 \, \mathrm{M}_{\odot}$ for the donor star. The WD is at least the third most massive ever found in a CV. Table~\ref{tab:parameters_summary} compiles all the fundamental parameters of XY~Ari presented in this work.

We compared our dynamical WD mass with estimates from X-ray spectral modelling values and found disagreement. Specifically, we observe that even recent models applied to hard X-ray ($>20$~keV) data may underestimate the WD mass. We also made a comparison with estimates from X-ray iron emission-line ratios, but were unable to find any clear trend due to the large uncertainties reported and the inconsistency between the works that have used this method. 

The present work is another piece of evidence that WD masses from current X-ray methods may suffer from significant systematic deviations. Further testing of these techniques requires precise WD mass determinations in X-ray bright IPs. More accurate dynamical studies would serve very well to this purpose.

\section*{Acknowledgements} 
We thank the anonymous referee for insightful comments and constructive suggestions. We are thankful to the GTC staff, in particular Antonio L. Cabrera Lavers, for their help and assistance in obtaining the spectroscopy presented in this paper. We thank Alina Streblyanska for sharing her expertise in EMIR with us. We also thank Nicol\'as Cardiel L\'opez for clarifying us some aspects about \textsc{pyemir}. This work was supported by the Agencia Estatal de Investigaci\'on del Ministerio de Ciencia e Innovaci\'on (MCIN/AEI) and the European Regional Development Fund (ERDF) under grants AYA2017--83216--P, AYA2017--83383--P and PID2021--124879NB--I00. MAPT was also supported by a Ram\'on y Cajal Fellowship RYC--2015--17854. PR-G acknowledges support from the Consejería de Economía, Conocimiento y Empleo del Gobierno de Canarias and the European Regional Development Fund (ERDF) under grant with reference ProID2021010132. The use of the \textsc{molly} package developed by Tom Marsh is acknowledged.

\section*{Data availability} 
The data used in this article are publicly available at the 10.4-m Gran Telescopio Canarias archive (\url{https://gtc.sdc.cab.inta-csic.es/gtc/}) under program ids GTC76-17B, GTC152-18B, GTC108-20B and GTCMULTIPLE2E-21A.



\bibliographystyle{mnras} 
\bibliography{bibliography} 


\bsp	
\label{lastpage}
\end{document}